\documentstyle[12pt]{article}
\newcommand{\be}{\begin{equation}}
\newcommand{\ee}{\end{equation}}
\newcommand{\bea}{\begin{array}}
\newcommand{\eea}{\end{array}}

\title{SOME REMARKS ON QUANTUM THEORY AND INTEGRABLE SYSTEMS}
\author{Robert Carroll\\University of Illinois\\Urbana, IL 61801
\thanks{email:  rcarroll@symcom.math.uiuc.edu}}

\date{July, 1996}

\begin{document}

\bibliographystyle{plain}
\maketitle

\begin{abstract}
Some formulas and speculations are presented relative to integrable
systems and quantum mechanics.
\end{abstract}


\section{INTRODUCTION}
\renewcommand{\theequation}{1.\arabic{equation}}\setcounter{equation}{0}

The point of departure here is an important paper \cite{fa}
which connects the probability density
$|\psi|^2$ with the space coordinate and a prepotential ${\cal F}$ generated
by duality ideas related to Seiberg-Witten (SW) theory (cf. \cite{sa}).
In fact a duality between the space coordinate and the wave function $\psi$
is established.  We will inject some ideas from integrable system theory
and quasiclassical analysis into this formalism in several ways and
thereby establish an heuristic connection with the Kadomtsev-Petviashvili
(KP) hierarchy.  The ensuing formulas suggest 
(in a conjectural and speculative spirit)
a number of possibilities
for further development involving integrability ideas and their multiple
roles in quantum theory.  

\section{BACKGROUND}
\renewcommand{\theequation}{2.\arabic{equation}}\setcounter{equation}{0}

In this section we will write down some formulas from \cite{fa} which
follow from direct calculation, without going into the philosophy of
\cite{fa}.  Thus one considers
\be
-\frac{\hbar^2}{2m}\psi''+V(X)\psi=E\psi
\label{A}
\ee
where we use $X$ as the quantum mechanical (QM) space coordinate with
$\psi'\sim\partial\psi/\partial X$. Before exhibiting the relevant
equations from \cite{fa} let us make a remark about the connections to
KP which are envisioned here.
\\[3mm]\indent {\bf REMARK 2.1.}$\,\,$   One connection we will want to
develop later involves writing e.g. $\epsilon=\hbar/\sqrt{2m}$ and
$\epsilon x=X$ with $\partial_x=\epsilon
\partial_X$ so that (\ref{A}) becomes
\be
\psi_{xx}-v(x)\psi=-E\psi=-\lambda^2\psi
\label{B}
\ee
provided one can write e.g.
$v(X/\epsilon)=V(X)+O(\epsilon)$ for some function $v$.  Then (\ref{A})
could be thought of formally
as $L^2_{+}\psi=\mu\psi$ for some Lax operator
$L=\partial +\sum_1^{\infty}u_i\partial^{-i}$.
We recall that in standard transitions from KP to dispersionless KP
(dKP) one begins with $L$ and supposes e.g. $u_i(X/\epsilon,T/\epsilon)
=U(X,T)+O(\epsilon)$ with $\psi=exp[(1/\epsilon)S(X,T,\mu)]$ (cf. \cite
{ca,cb,ta} - we are taking $T$ here as $T_2,T_3,\cdots)$ or simply as $T_2$).  
Then the equations of KP lead (formally - algebraicaly) to
the dKP hierarchy.  In the present situation an equation $L^2_{+}\psi=
\psi_{xx}-v(x)\psi=\mu\psi$ could correspond to $\epsilon^2\psi''+O(\epsilon)
\psi -(V(X)+O(\epsilon))\psi$ however, which for $\psi=exp[(1/\epsilon)S]$
becomes, with $S'=P,\,\,(P^2-V)\psi+O(\epsilon)\psi=\mu\psi\to (P^2-V)=\mu$
(note $\epsilon^2\psi''\sim \epsilon(S'exp(S/\epsilon)'=(\epsilon S''+
P^2)exp(S/\epsilon)$).  Thus, in certain situations, 
one could think of $\epsilon^2
\psi''-V(X)\psi=\mu\psi$ as a QM problem related to a KP situation
$\psi_{xx}-v(x)\psi=\mu\psi$.  The unwieldy expression $\psi=exp[(1/
\epsilon)S]$ in KP theory becomes a natural QM wave function with 
small parameter $\hbar$.  The passage from $v(x)$ to $V(X)$ is intricate
and technical and we refer to \cite{la} for details; there will be
in any case realistic situations where this correspondence makes sense.
With this in mind one could now consider
\be
i\hbar\psi_{\tau}=-\frac{\hbar^2}{2m}\psi''+V\psi
\label{C}
\ee
Scale $\tau\to T=-(1/\sqrt{2m})\tau$ for example so $\psi_{\tau}=(1/
\sqrt{2m})\psi_T$ and (\ref{C}) becomes $-i\epsilon\psi_T=-\epsilon^2\psi''
+V\psi\,\,(\psi'\sim\partial\psi/\partial X)$.  Then for $\epsilon x=X$ and
$i\epsilon t=T$ one has
\be
\psi_t=\psi_{xx}-v\psi=L_2\psi
\label{D}
\ee
This is now in KP form with $t$ playing the role of $y\sim t_2$; more
precisely take $\epsilon t=T\sim T_2$ and $y=it$ so 
$\partial_t=\epsilon\partial_T$
with $\partial_y=i\partial_t$ which implies $\partial_t=i\epsilon\partial_T$
and $\partial_y\psi=L_2\psi$.  Here we recall that there are two forms of
KP, namely KP1 and KP2, where in KP1 one takes the even variables $t_{2n}$ to
be imaginary in many situations.  Thus $t\sim t_2$ imaginary is quite natural.
We note that for a general correspondence between QM and KP we are most
emphatically not generally in a KdV situation since $L_2=L^2_{+}=L^2
\Rightarrow\partial_yL=[L^2,L]=0$ with no $y$ dynamics.  
In order to get equations such as (\ref{A}) 
one can think of $i\hbar\psi_{\tau}=-(\hbar^2/2m)\psi''+V\psi={\cal H}\psi$
and if ${\cal H}$ is independent of $\tau$ (i.e. $V=V(X)$) then $\psi=
exp(\tau{\cal H}/i\hbar)\psi_0(X)$ involves $i\hbar\psi_{\tau}={\cal H}
\psi$ which is a dynamical version of
(\ref{A}).  This suggests that the study of KP with $u_1$ 
independent of $t_2$ should be of special interest.  Also since the theory
of algebraic curves is closely associated to KP (cf. \cite{cc} for 
information on the Schottky problem, etc.) one has
yet another natural entry point of Riemann surfaces into QM via the 
correspondence indicated.  
A further possibility is to note that the identity
$\bar{\psi}\sim\psi^D$ in \cite{fa} could be extended to $\psi^*\sim\psi^D$
where $\psi^D$ now refers to duality in the QM sense while $\psi^*\sim
\psi^{\dagger}$ denotes KP duality.  Note here that for $V$ real one can
write ($B_2\sim L^2_{+}$)
\be
\partial_y\psi=(\partial_x^2-v)\psi=-H\psi=B_2\psi;\,\,-\partial_y\psi^*
=B_2^*\psi=B_2\psi^*=-H\psi^*
\label{E}
\ee
But for $y=it,\,\,\psi=exp(-itH)\psi_0=exp(-yH)\psi_0$ implies $\bar{\psi}=exp
(itH)\bar{\psi}_0=exp(yH)\bar{\psi}_0$ so $\bar{\psi}\sim\psi^*=exp(yH)
\psi^*_0$ for $\psi^*=\bar{\psi}_0$.  We will use $\bar{\psi}$ for $\psi^D$
below following \cite{fa} and this suggests that (perhaps in a limited
way only) one can relate $\psi^*$ to $\psi^D$ and envision QM duality as
related to a $\psi - \psi^*$ duality.  This will be discussed more below.
$(\bullet)$
\\[3mm]\indent
Now in \cite{fa} one establishes a number of equations following from
(\ref{A}) by introducing a prepotential ${\cal F}$ such that $\partial
{\cal F}/\partial\psi=\psi^D$ where $\psi^D$ may be identified with $\bar
{\psi}$ in generic situations.  The philosophy of such equations is related to
ideas emerging from electromagnetic (EM) duality as in \cite{ba,bb,cd,
ea,ia,kc,kb,ma,na}.  One can also derive such equations by direct calculation
and we list here a few of these useful relations without discussing
the philosophy.  The Wronskian in (\ref{A}) is taken to be
$W=\psi'\bar{\psi}-\psi\bar{\psi}'=2\sqrt{2m}/i\hbar=2/i\epsilon$ and one
has ($\psi=\psi(X)$ and $X=X(\psi)$ with $X_{\psi}=\partial X/\partial\psi=
1/\psi'$)
\be
{\cal F}'=\psi'\bar{\psi};\,\,{\cal F}=\frac{1}{2}\psi\bar{\psi}+\frac
{X}{i\epsilon};\,\,
\frac{\partial\bar{\psi}}{\partial\psi}=\frac{1}{\psi}\left[\bar{\psi}-
\frac{2}{i\epsilon}X_{\psi}\right]
\label{F}
\ee
Setting $\phi=\partial{\cal F}/\partial(\psi^2)=\bar{\psi}/2\psi$ with
$\partial_{\psi}=2\psi\partial/\partial(\psi^2)$ and $\partial\phi/\partial
\psi=-(\bar{\psi}/2\psi^2)+(1/2\psi)(\partial\bar{\psi}/\partial\psi)$
one has a Legendre transform pair
\be
-\frac{X}{i\epsilon}=\psi^2\frac{\partial{\cal F}}{\partial(\psi^2)}
-{\cal F};\,\,-{\cal F}=\phi\frac{1}{i\epsilon}X_{\phi}-\frac{X}{i\epsilon}
\label{H}
\ee
One obtains also
\be
|\psi|^2=2{\cal F}-\frac{2X}{i\epsilon}\,\,({\cal F}_{\psi}=\bar{\psi});\,\,
-\frac{X}{i\epsilon}X_{\phi}=\psi^2;\,\,{\cal F}_{\psi\psi}=\frac
{\partial\bar{\psi}}{\partial\psi}
\label{I}
\ee
Further from $X_{\psi}\psi'=1$ one has $X_{\psi\psi}\psi'+X^2_{\psi}\psi''
=0$ which implies
\be
X_{\psi\psi}=-\frac{\psi''}{(\psi')^3}=\frac{1}{\epsilon^3}\frac
{(E-V)\psi}{(\psi')^3}
\label{K}
\ee
\be
{\cal F}_{\psi\psi\psi}=\frac{E-V}{4}({\cal F}_{\psi}-\psi\partial^3_{\psi}
{\cal F})^3=\frac{E-V}{4}\left(\frac{2X_{\psi}}{i\epsilon}\right)^3
\label{L}
\ee
\section{WKB AND QUASICLASSICAL THEORY}
\renewcommand{\theequation}{3.\arabic{equation}}\setcounter{equation}{0}

There are (at least) two ways to inject quasiclassical ideas into the
picture.  First we indicate an eikonal transformation d'apr\`es \cite{mb}.
Thus write $p=\Im\psi$ and $q=\Re\psi$ with
\be
H_1=\frac{1}{2}\int dX\left[\frac{\hbar^2}{2m}\{(p')^2+(q')^2\}+V(p^2+q^2)
\right]
\label{M}
\ee
Then formally
\be
\frac{d}{dt}{p \choose q}=\frac{1}{\hbar}
\left(
\begin{array}{cc}
0 & -1\\
1 & 0
\end{array}\right)
\left(\begin{array}{c}
\frac{\delta H_1}{\delta p}\\
\frac{\delta H_1}{\delta q}
\end{array}
\right)
\label{N}
\ee
Set now for real $A$ and $S$
\be
\psi=Ae^{(i/\epsilon)S};\,\,p=ASin(\frac{1}{\hbar}S);\,\,q=ACos(\frac{1}{\hbar}
S)
\label{O}
\ee
Introduce new variables 
\be
\chi=A^2=|\psi|^2;\,\,\xi=\frac{1}{2\hbar}S
\label{P}
\ee
and set ($'\sim\partial/\partial X$)
\be
K_1=\int dX\left[\frac{\hbar^2}{2m}\left(\frac{(\chi')^2}{4\chi}+2\chi
(\xi')^2\right)+V\chi\right]
\label{Q}
\ee
Then it follows that
\be
\dot{\xi}=\frac{\hbar}{2m}\frac{(\sqrt{\chi})''}{\sqrt{\chi}}-\frac{\hbar}{m}
(\xi')^2-\frac{1}{\hbar}V;\,\,\dot{\chi}=-\frac{2\hbar}{m}(\chi\xi')';\,\,
\delta p\wedge \delta q=\delta\xi\wedge\delta\chi=\tilde{\omega}
\label{R}
\ee
Thus formally one has a Hamiltonian format with symplectic form as in
(\ref{R}).
\\[3mm]\indent
It is interesting to write down the connection between the $(S,A)$ or
$(\chi,\xi)$ variables and the variables of Section 2 from \cite{fa}.  Thus
\be
{\cal F}=\frac{1}{2}\chi+\frac{X}{i\epsilon};\,\,{\cal F}'=\frac{1}{2}\chi'
+\frac{i}{\epsilon}P\chi
\label{T}
\ee
for $S'=S_X=P$ and there is an interesting relation
\be
P\chi=-1\Rightarrow \delta\chi=-\frac{\chi}{P}\delta P
\label{U}
\ee
Further from $\phi=(1/2)exp[-(2i/\epsilon)S]$ and $\psi^2=\chi exp(4i\xi)$ 
we have
\be
\psi^2\phi=\frac{1}{2}\chi=-\frac{1}{\epsilon}\phi X_{\phi};\,\,\xi=\frac
{S}{2\epsilon}=\frac{i}{4}log(2\phi)
\label{V}
\ee
Now the theory of the Seiberg-Witten (SW) differential $\lambda_{SW}$ 
following \cite{baa,cd,da,ea,ia,ka,kb,ma,md,na,sa} 
involves finding a differential
$\lambda_{SW}$ of the form $QdE$ or $td\omega_0$ (in the spirit of
\cite{kb} or \cite{da,ia} respectively) such that $d\lambda_{SW}=\omega$
is a symplectic form (cf. \cite{cd,ia,md} for some discussion).  In the
present context one can ask now whether the form $\tilde{\omega}$ of 
(\ref{S}) makes any sense in such a context.  Evidently this is jumping
the gun since there is no Riemann surface in sight (but see Sections 4
and 5); the motivation to consider the matter here comes from the
following formulas which express $\tilde{\omega}$ nicely in terms of
the duality variables of Section 2.
Thus a priori $\psi=\Re\psi+i\Im\psi$ has two components which are also
visible in $\psi=Aexp(iS/\epsilon)$ as $A$ and $S$.  The relation $P\chi
=\chi(\partial S/\partial X)=-1$ indicates a dependence between $A$
and $S'$ (but not $A$ and $S$) which is a consequence of the duality between
$\psi$ and $X$.  Then $2AS'\delta A+A^2\delta S'=0$ or $\delta S'
=-(2S'/A)\delta A\equiv (\delta S'/S')=-2(\delta A/A)$, whereas $\delta
\psi/\psi\sim 2(\delta A/A) +(i/\epsilon)\delta S$.  It follows that
$\Re(\delta\psi/\psi)=-(\delta S'/S')$ and $\Im(\delta \psi/\psi)=(\delta
S/\epsilon)$.  The sensible thing seems to be to look at the complex
dependence of $X(\psi)$ and $\psi(X)$ in terms of two real variables
and $\delta\xi\wedge\delta\chi$ will have a nice
form in transforming to the variables of \cite{fa}.  In particular
from $\psi^2\phi=(1/2)\chi$ with $\delta\chi=4\phi\psi\delta\psi+
2\psi^2\delta\phi$ we obtain
$(\delta\psi/\psi)=2(\delta\chi/\chi)-
(\delta\phi/\phi)$.
Hence one can write 
\be
\delta\xi\wedge\delta\chi=\frac{i}{4}\frac{\delta\phi}{\phi}
\wedge\chi\frac{\delta\chi}{\chi}=\frac{i}{2}\delta\phi\wedge\delta\psi^2
\sim\frac{i}{2}\delta\bar{\psi}\wedge\delta\psi
\label{X}
\ee
(note $\delta\phi=(1/2\phi)\delta\bar{\psi}-(\bar{\psi}/2\psi^2)\delta\psi$)
and in an exploratory spirit the differentials $\lambda=(i/2)\phi\delta
\psi^2$ or $\lambda=(i/2)\psi^2\delta\phi$, along
with $\lambda=(i/2)\bar{\psi}\delta\psi$
or $\lambda=(i/2)\psi\delta\bar{\psi}$, might merit further consideration.

\section{CONNECTIONS TO KP}
\renewcommand{\theequation}{4.\arabic{equation}}\setcounter{equation}{0}

We refer now to \cite{ca,cb,ta} for dispersionless KP and consider
$\psi=exp[(1/\epsilon)S(X,T,\lambda)]$ instead of $\psi=Aexp(S/\epsilon)$.
Thus $P=S'=S_X$ and $P^2=V-\lambda^2$ (for $E=\lambda^2$).  One computes
easily
(recall $X_{\psi}=1/\psi'$ and $\psi'=(P/\epsilon)\psi$)
\be
\phi=\frac{1}{2}e^{-(2i/\epsilon)\Im S};\,\,\frac{1}{\epsilon}X_{\phi}=
-ie^{(2/\epsilon)S};\,\,X_{\psi}=\frac{\epsilon}{P}e^{-S/\epsilon}
\label{AA}
\ee
\be
\frac{1}{\epsilon}X_{\psi\psi}=\frac{E-V}{P^3}e^{-S/\epsilon};\,\,
{\cal F}_{\psi}=\bar{\psi}=e^{\bar{S}/\epsilon};\,\,
{\cal F}_{\psi\psi}=e^{-(2i/\epsilon)\Im S}-\frac{2}{iP}e^{-2S/\epsilon}
\label{AB}
\ee
Next from ${\cal F}'=\psi'\bar{\psi}=(P/\epsilon)exp[(2/\epsilon)\Re S]$
and $W=(2/i\epsilon)=(\psi'\bar{\psi}-\psi\bar{\psi}')=({\cal F}'
-\bar{{\cal F}}')$ we have $\Im{\cal F}'=-(1/\epsilon)$ and
from ${\cal F}=(1/2)\psi\bar{\psi}+(X/i\epsilon)$ we see that $\Im {\cal F}
=-(X/\epsilon)$.  In addition
\be
|\psi|^2=e^{(2/\epsilon)\Re S};\,\,\frac{S}{\epsilon}=\frac{1}{2}log|\psi|^2
-\frac{1}{2}log(2\phi)
\label{AC}
\ee
Finally from $\partial\bar{S}/\partial S=(\psi/\bar{\psi})(\partial\bar{\psi}/
\partial\psi)=1-(2/iP)exp[-(2/\epsilon)\Re S]$ one has
\be
\bar{S}_X=\bar{P}=P-\frac{2}{i}e^{-(2/\epsilon)\Re S};\,\,\frac{X}{i\epsilon}
={\cal F}-\frac{1}{2}e^{(2/\epsilon)\Re S};\,\,e^{-(2/\epsilon)\Re S}=
\frac{i}{2}(P-\bar{P})
\label{AD}
\ee
Now from \cite{cb} for example one can write
\be
P-\bar{P}=\lambda-\bar{\lambda}-\sum_1^{\infty}\frac{1}{j}\left(F_{1j}
\lambda^{-j}-\bar{F}_{1j}\bar{\lambda}^{-j}\right)
\label{ADD}
\ee
where $F$ is the logarithm of the quasiclassical or dispersionless tau
function.  If one thinks here of a KdV situation, $F_{1j}=0$ except for
$j=2n-1$, but we recall that KP was the more interesting habitat here
(cf. Remark 2.1).  One expects the $F_{1j}$ above to be real so for
$\lambda=Rexp(i\theta)$ we have
\be
P-\bar{P}=\lambda\left(1-e^{-2i\theta}\right)-\sum_1^{\infty}\frac{F_{1m}}{m}
\left(1-e^{2im\theta}\right)\lambda^{-m}=
\lambda A_0-\sum_1^{\infty}A_m\lambda^{-m}
\label{AE}
\ee
where $A_m=A_m(\theta)$.  It follows easily that
\be
\frac{1}{P-\bar{P}}=\frac{1}{\lambda A_0}\left[\sum_0^{\infty}\left(\sum_1^
{\infty}\frac{A_m}{A_0}\lambda^{-m-1}\right)^n\right]=\sum_1^{\infty}c_p
(\theta)\lambda^{-p}
\label{AF}
\ee
where the $c_p$ can be computed recursively from the $A_m$.  Then from
(\ref{D}) one has
\be
\Im{\cal F}=-\frac{X}{\epsilon};\,\,\Re{\cal F}=\frac{-1}{2\Im P}=
\frac{-i}{P-\bar{P}}=-i\sum_1^{\infty}c_p\lambda^{-p}
\label{AG}
\ee
from which
\be
\widetilde{Res}({\cal F}\lambda^{p-1})=-ic_p(\theta)
\label{AH}
\ee
where $\widetilde{Res}$ is a formal power series 
residue (no contour integration).
In the present situation $|\psi|^2=exp[(2/\epsilon)\Re S]$ and $2\phi=
exp[-(2i/\epsilon)\Im S]$ can play the roles of independent variables
(cf. (\ref{AC}).  The version here of $P\chi=-1$ is $\chi\Im P=-1$,
while $\psi^2\phi=(1/2)|\psi|^2=(1/2)\chi$ again, and
we obtain as in Section 3 the formula (\ref{X}).  Let us note also
from (\ref{AD}) that 
\be
\partial_{\lambda}{\cal F}=\frac{2}{\epsilon}\left({\cal F}-\frac{X}{i\epsilon}
\right)\Re S_{\lambda}\Rightarrow\partial_{\lambda}log({\cal F}-\frac
{X}{i\epsilon})=\frac{2}{\epsilon}\Re{\cal M}
\label{AI}
\ee
where ${\cal M}$ is the dispersionless Orlov-Schulman operator (cf. \cite
{ca,cf,ta}).

\section{REMARKS}
\renewcommand{\theequation}{5.\arabic{equation}}\setcounter{equation}{0}

Seiberg-Witten differentials require a Riemann surface, period integrals,
monodromy, moduli, etc. and it is not clear what kind of abstract definition
(if any) would be realistic.  One thinks in general of an ``action" 
form $\sum p_idq_i\sim\theta$ for example with $(q,p)\in T^*M$ and
$TM - T^*M$ connected by the canonical Legendre transform with $d\theta
\sim\omega=\sum dp_i\wedge dq_i$.  Given a duality as introduced in \cite{fa}
one asks whether the formalism extends in some way ``abstractly" to a
natural geometrical object based perhaps on $(p,q)$ as in Section 4, or
on $(A,S),\,\,(\Re S,\Im S)$, or $(\phi,\psi^2)$.  One could also envision
taking e.g. expectation values of such objects also to create a ``numerical"
manifold.
\\[3mm]\indent
Now the introduction of KP or KdV ideas here does give us a Riemann surface
to attach to the framework.  One point of view would suggest looking at a
Riemann surface based on $P^2=V-\lambda^2$ in a KdV context (cf. \cite{cb})
with $P=\pm\sqrt{V-\lambda^2}$ and a cut between $(-\sqrt{V},\sqrt{V})$
in the $\lambda$ plane.  Another (more attractive)
point of view involves looking at a finite
zone KP or KdV situation.  In particular for KdV with a periodic finite
zone potential one will have not only a hyperelliptic surface but also some
monodromy information via the Floquet theory (cf. \cite{mc}).  Some 
caution is necessary since a potential $v\sim u_1$ 
at the second ($L^2_{+}$) level
may give a hyperelliptic KdV situation when $L^2_{+}=L^2$ but may or may
not generate finite zone situations for general corresponding KP theories.
We note that $u=u_1$ determines all $u_n$ in a corresponding Lax operator
for KP modulo $\partial^{-1}$ (cf. \cite{cf}); this means many possible
operators $L$ could be associated with $u_1$ and it might be of interest
to investigate whether any of them are finite zone for example (beyond
the KdV extension).  The QM equation (\ref{A}) corresponds to the 
genus zero version of the slow
variable equation for a modulated wave train with $\psi$ itself arising
from (\ref{B}).  Then for (\ref{B})
one has averaging procedures for finite zone
situations as in \cite{cd,ce,fb,fc,ka}.  On the associated Riemann surface
$\Sigma_g$ (of genus $g$) there are 
(for hyperelliptic KdV situations) $2g-1$ independent branch points 
$\Lambda_i$ which serve as moduli.  In general there will be $3g-3$ moduli
on a Riemann surface $\Sigma_g$ and for any Riemann surface
one can generate the Whitham dynamics,
differential forms, Baker-Akhiezer (BA) functions, etc. following
\cite{cd,ce,fb,fc,ia,ka,kb}.  One can produce many situations involving
differentials $dQ$ and $dE$ such that action integrals $a_i=\oint_{A_i}QdE$ 
are adapted to problems in EM duality (cf. \cite{cd,kb,md}).
The $E$ here is not a priori related to the $E$ of Section 2 and we note 
that perhaps different normalizations of differentials (e.g. $\Im\oint_{A_i}
d\Omega_n=\Im\oint_{B_i}d\Omega_n=0$) should be used in order to best
fit the theory of \cite{kb} to the QM situation as in Section 2.  
In particular $a_i=\oint_{A_i}pdE$  
is naturally adapted to a KdV situation relative to the Gardner-Faddeev-
Zakharov symplectic bracket
\be
\omega_{GFZ}=-\frac{1}{2}<\delta v\wedge\partial^{-1}\delta v>
\label{BA}
\ee
where $<\,\,\,>$ denotes ergodic averaging. 
Note that the symplectic forms obtained in Sections 3 and 4 are based
on a genus zero situation corresponding to the dispersionless theory
and WKB analysis so a priori one should be careful about
expecting that
(\ref{X}) for example will carry over here; even though the forms are
very ``canonical" in appearance they may describe a different structure
from (\ref{BA}).
Equation (\ref{BB}) below
does give an 
expression for a symplectic form based on the dual variables $\psi$
and $\bar{\psi}\sim\psi^*$, which reduces to (\ref{BA}), but it does not
seem to reduce to (\ref{X}).  Thus it appears that (\ref{BA}) - (\ref{BB})
and (\ref{X}) do indeed refer to different structures and for a SW 
differential one should use $\lambda=pd\Omega_2$ with $\omega=d\lambda$
given by (\ref{BA}) - (\ref{BB}).  In any event
the Riemann surface should be related to the problem
via $v$ as indicated in the discussion which follows.
Here $v$ is expressed in terms of theta
functions in the fast variables $x,y\sim it$ with the moduli ($\Lambda_i$
or more generally some $h_i$) being functions of the slow variables
$X,Y\sim T$.  The averaging kills the fast variables.  The genus zero
theory should be directly associated to the dispersionless limit theory
(cf. \cite{ca,cb,ta}) while the more interesting situation where $g>1$
will involve Whitham theory as in \cite{cd,ce,fb,fc,ia,ka,kb}.  Thus one
thinks of averaging with the $\psi$ of (\ref{B}) where $\psi^*\sim\bar{\psi}$
as in Remark 2.1.  The Riemann surface here could then be built from a KP 
finite zone problem and the equation $\partial_y\psi=-H\psi=E\psi$
corresponds to an eigenmode of $H=-\partial^2+V$ 
(recall $t_2\sim y$ corresponding to $it$ where
$\epsilon t=T$ is the QM time - cf. Remark 2.1).
\\[3mm]\indent
Now we still have the question
of what to look at here in terms of a Riemann surface.
There are a number of apparently open questions.  Thus let $L$ correspond to
a finite zone KP situation; then we have a Riemann surface $\Sigma_g$ and
any discussion of a symplectic form $\omega_{{\cal M}}$
as in \cite{cd,kb}, built upon
$L_2\sim L^2_{+}$ as in \cite{cd,kb,md}, will be natural and appropriate.
This is probably the point of view to take here.  If in fact $v$ generates
a finite zone KdV situation with $L^2_{+}=L^2$ (with no extrapolation to
KP beyond the natural KdV hierarchy extension) then the $y\sim it$ dynamics is
not present in the QM analogue and one is dealing with a strictly stationary
situation (based on a KdV surface $\Sigma_g$ now).  In any event $y$ does
not appear in $\omega_{{\cal M}}$ so both situations would be meaningful.
\\[3mm]\indent
The point now is that given a Riemann surface attached ``naturally" to the 
QM problem (\ref{A}) and its associated duality theory, it will be possible
to discuss SW differentials and symplectic forms in the spirit of \cite{cd,kb}.
Thus briefly one has differentials $d\Omega_n$ associated to ``Lax" operators
$L_n\sim L^n_{+}$ and $\lambda_{SW}\sim pdE$ for example with $a_i=\oint_{A_i}
\lambda_{SW}$.  The natural correspondence to \cite{cd,kb} here is to take
$dE=d\Omega_2$ with associated (KdV type) symplectic form $\omega\sim
\omega_{{\cal M}}$ as in (\ref{BA}), with associated versions
$$
\omega_{{\cal M}}=\delta\left(\sum_1^gp(\gamma_i)dE(\gamma_i)\right)=
\sum_1^{\infty}\delta p(\gamma_i)\wedge dE(\gamma_i)=$$
\be
=\sum_1^gda_i\wedge d\omega_i
=-Res_{\infty}\frac{<\delta\psi^*\wedge\delta L_2\psi>}{<\psi^*\psi>}dp
\label{BB}
\ee
The attachment of $\Sigma_g$ to the QM problem and its associated
integrable system arises then through $L_2$, plus the dependence of
differentials, etc. on the surface ``generated" by $L$ or $L_2$.
The $v - V$ correspondence is thus critical.
\\[3mm]\indent {\bf Acknowledgement}.$\,\,$  The author would like to
thank M. Matone for helpful comments on \cite{fa} and Y. Nutku for useful
discussion on related matters.

\end{document}